\documentclass[useAMS,usenatbib]{mn2e}
\usepackage{amsmath}
\usepackage{graphicx}
\usepackage{color}
\usepackage{ulem}

\usepackage{epsf}

\title[HLX--1 may be an SS433 system] {HLX--1 may be an SS433 system}

\author[Andrew King and Jean--Pierre. Lasota]
{Andrew King$^{1, 2}$ and Jean--Pierre Lasota$^{3,4,5}$\\
$^{1}$ Theoretical Astrophysics Group, Department of Physics \&Astronomy, University of Leicester, Leicester LE1 7RH, UK\\
$^{2}$ Astronomical Institute Anton Pannekoek, University of Amsterdam, Science Park 904, 1098 XH Amsterdam, Netherlands\\
$^{3}$ CNRS, UMR 7095, Institut d'Astrophysique de Paris, 98bis Bd Arago, 75014 Paris, France\\
$^{4}$ Nicolaus Copernicus Astronomical Center, Bartycka 18, Warsaw, Poland\\
$^{5}$ Sorbonne Universit\'es, UPMC Univ Paris 06, UMR 7095, 98bis Bd Arago, 75014 Paris, France      
}

\date{\today}

\volume{000}

\pagerange{\pageref{firstpage}--\pageref{lastpage}} \pubyear{2014}

\begin{document}

\label{firstpage}

\maketitle

\begin{abstract}
We show that the hyperluminous source HLX--1 may be a stellar--mass binary system like SS433, but seen along its X--ray beams. The precession of these beams gives the $\sim 1$~yr characteristic timescale of the light curve, while the significant X--ray duty cycle means that the precession angle must be comparable with the beam opening angle, which is of order $1.6^{\circ}$. The X--ray light curve appears to result from geometric collimation and scattering as the beam moves through the line of sight. Encouragingly, the distance $\sim 95$~Mpc suggested for HLX--1 is only a few times larger than the minimum distance at which we can expect to view such a highly--beamed system along its axis. This picture allows a simple interpretation of HLX--1 as the most extreme known member of the ULX population.
\end{abstract}


\section{Introduction}

The X--ray source 2XMM J011028.1--460421 (better known as HLX--1) \citep{Farrell09}
has attracted attention because it is positionally coincident with
the outer regions of the edge--on spiral galaxy ESO 243--49 \citep{Farrell09} at a redshift of 0.0224. The detection at the position of HLX--1 of a narrow $H_\alpha$ emission line with redshift corresponding to that of ESO 243--49 \citep{Wiersema10,Soria13} has been seen as confirming the association of HLX--1 with this galaxy,  although the association of this line with the accretion flow emitting the X--rays has still to be conclusively demonstrated.

The assumption that HLX--1 is physically associated with this galaxy (at a distance $D = 95$~Mpc)  has far--reaching consequences. It implies an unabsorbed isotropic 0.2--10~keV luminosity for HLX--1 of $L_{\rm max} = 1.3\times 10^{42}$~erg\,s$^{-1}$ at maximum, making it the brightest known hyperluminous X--ray source (HLX).  The additional assumptions that the source radiates at no more than ten times the standard Eddington luminosity \citep{Begelman02, Begelman06}, and that its emission is isotropic, would give a minimum accretor mass $\sim 500~\mathrm{M}_\odot$ In this sense HLX--1 is the best current candidate for an intermediate--mass black hole (IMBH). 

Observations of HLX--1 place tight constraints on models. The source shows a sequence of near--regular outbursts lasting $\sim 200$ days. For the four outbursts between 2009 and 2012 the recurrence time was $\sim 370$ days, but the 2013 outburst started about 1 month `late' \citep{Godet13a}. Multiwavelength observations of HLX--1 during 2009--2013 reveal outburst properties resembling those of low--mass X--ray binaries  (LMXBs: e.g. Remillard \& McClintock 2006) in several respects. It is widely accepted that these outbursts are well described by the thermal--viscous disc instability model \citep[TVDIM,][]{Lasota01,Dubus01}, when account is taken of self--irradiation of the disc by the central X--rays \citep[cf][]{KingRitter98}. However
\citet{Lasota11} showed that a disc instability model with an IMBH accretor could not explain the light curve of HLX--1 if the system was assumed to be at the 95~Mpc distance implied by the putative association with ESO 243--49. In fact all such models require the source to be well within the Local Group, and to have a stellar mass accretor  \citep{Lasotaetal14}. 
 
Given this difficulty, more unusual models have appeared. These inevitably pay the price of requiring very special conditions. \citet{Lasota11} suggested that the outbursts might be periodic mass transfer events on to an IMBH accretor, triggered when a star on an eccentric orbit about the black hole fills its tidal lobe at  pericentre.
The problem here is that the $\sim 0.5\,{\rm yr}$ decay time of the light curve is presumably viscous, and so requires a small disc radius $R\sim 10^{11}\,{\rm cm}$ \citep{Lasota11}. But to give the nearly periodic repetitions  of the outbursts, the orbital period must be of order 1~yr, implying a semi--major axis $a \simeq 3\times 10^{14}\,{\rm cm}$, and so an orbital eccentricity $e$ improbably close to unity ($(1 - e) \simeq R/a\simeq 3\times 10^{-4}$). There are problems with the stability of such an orbit, but worse, the 1--month `delay' of the 2013 outburst is very hard to reconcile with the eccentric binary picture. Further, the observed outburst rise times of only a few days appear to require an accretion disc structure very different from anything so far considered \citep{Lasota11,Webb13}. 
\citet{Miller14} proposed a modified version of the IMBH--enhanced mass-transfer model in which a high-mass giant star has had most of its envelope tidally stripped by an IMBH in HLX--1 only $\sim 10$ years ago. The remaining core plus low--mass hydrogen envelope is assumed to be currently feeding the IMBH via an (unrealistically) strong wind. This model produces a disc smaller than in the Roche--lobe overflow case but the rise--time and X/V delay still require non--standard disc physics. In addition, we must be viewing this system at an extremely unusual epoch.
 
But there is a simple alternative model compatible with the 95 Mpc distance. We take HLX--1 as the brightest known ultraluminous X--ray source (ULX). A commonly--accepted model for ULXs is that they are stellar--mass X--ray binaries with such high mass transfer rates that this leads to geometrical collimation or beaming of most of their emission \citep{Kingetal01}. In this picture the emitted fluxes are highly anisotropic, and multiplying them by $4\pi D^2$, where $D$ is the source distance, overestimates the intrinsic source luminosity by the inverse beaming factor $1/b \gg 1$. This picture is known to work well in explaining the majority of observed ULXs (see below), so it is worth asking if it can also explain HLX--1. We consider this idea here, and see that it leads to a suggestive analogy between HLX--1 and the well--known extreme Galactic binary SS433, often believed to be a ULX system viewed from outside the beaming angle.

\section{Beaming, Luminosity, Distance}

\citet{King09} used the observed scaling between soft X--ray luminosity and temperature in a class of ULXs to deduce the relation
\begin{equation}
b \sim {73\over \dot m^2}
\label{beam}
\end{equation}
between the beaming factor $b$ and the Eddington ratio $\dot m$, assumed $\ga 8$. We define the latter as 
\begin{equation}
\dot m = {0.1c^2\dot M\over L_{\rm Edd}},
\label{mdot}
\end{equation}
where $\dot M$ is the mass transfer rate feeding the accretion disc at large radii, $L_{\rm Edd}$ the Eddington luminosity of the mass--gaining compact object, and we have assumed an accretion efficiency $\eta = 0.1$. Although the form (\ref{beam}) is derived from considering a specific radiation component, the beaming represented by $b$ is assumed to be geometric, and to apply to photons of all energies emitted close to the accretor. \citet{King09} shows that the expression (\ref{beam}) agrees with theoretical expectations.

Combining (\ref{beam}) with the usual expression for the accretion luminosity at high Eddington ratios $\dot m \gg 1$ \citep{SS73} gives the apparent (i.e. assumed isotropic) luminosity of a ULX as
\begin{equation}
L_{\rm sph} = {1\over b}L_{\rm Edd}(1 + \ln\dot m) \simeq 1.8\times 10^{36}m_1\dot m^2(1 + \ln \dot m)~{\rm
  erg\, s^{-1}}
\label{lum}
\end{equation}
where $m_1 = M_1/M_{\odot}$ is the accretor mass in solar units (some authors insert dimensionless factors  of order unity, or the disc aspect ratio $H/R \sim 1$, in front of the $\ln \dot m$ term here. However the dominant scaling is in the $\dot m^2$ term expressing the beaming.) This expression implies a luminosity function for ULXs in good agreement with observations of ULXs in the Local Group  \citep{Mainierietal10}. The majority of ULXs have moderate
Eddington factors $\dot m \sim {\rm few}\times10$, and so $L_{\rm sph} \sim 10^{39} - 10^{41}~{\rm
  erg\, s^{-1}}$ (from \ref{lum}).

Applying (\ref{lum}) to HLX--1 ($L_{\rm sph} \simeq 1.3\times 10^{42}~{\rm erg\, s^{-1}}$) and assuming that the accretor is a $\simeq 10M_{\odot}$ black hole, gives an Eddington factor
\begin{equation}
\dot m \simeq 110. 
\label{edd}
\end{equation}
Since the Eddington accretion rate for this black hole mass is $\dot M_{\rm Edd} \simeq 10^{-7}M_{\odot}\,{\rm yr}^{-1}$ this implies a mass transfer rate 
$\dot M \simeq 10^{-5}M_{\odot}\,{\rm yr}^{-1}$ for HLX--1. This is similar to that inferred for SS433 \citep{Kingetal00, Begelman06}. Mass is transferred on the thermal timescale of the donor star, which is more massive than the compact accretor. This is a natural stage in the evolution of high--mass binaries, and directly follows the standard high--mass X--ray binary phase once the expanding donor star fills its Roche lobe.

\section{Distance}

The possible identification with a system like SS433 is attractive, but must pass several tests before we can accept it as a model for HLX--1. The first of these concerns the relation between beaming and distance. The Eddington factor (\ref{edd}) implies from (\ref{beam}) a beaming factor $b \simeq 6.0\times 10^{-3}$, and by elementary geometry, a beam opening angle $\theta_b \simeq 1.6^{\circ}$. We must first check that we do not require a cosmic conspiracy to be sitting in such a narrow beam, but instead that HLX--1 is at a sufficiently large distance, with so many similarly--beamed systems within this volume of space
that we would expect to be in the beam of at least one of them purely by chance. Specifically, we consider a population of similarly--beamed systems whose host galaxies have space density $n_g$~Mpc$^{-3}$. We assume that each host contains $N$ such systems,
with radiation beams oriented randomly. To be in the beam of one such
object one has to search through $\sim 1/Nb$ galaxies, i.e. a space
volume $\sim 1 /n_g N b$. The nearest suitably-oriented system of this type is thus at a
distance
\begin{equation}
D_{\rm min} \sim \left({3\over 4\pi n_g Nb}\right)^{1/3} \sim
13N^{-1/3}\dot m_{110}^{2/3}~{\rm Mpc}, 
\label{dmin}
\end{equation}
where $\dot m_{110} = \dot m/110 $, and we have assumed 
$n_g \sim 0.02~{\rm Mpc}^{-3}$, similar to L* galaxies, at the second step. (Note that the corresponding equation [15]) in \citet{King09} has the coefficient misprinted as 660 rather than 260.) We assume $N \sim 1$, i.e. that the number density $Nn_g$ of randomly--oriented systems with $\dot m \sim 110$ is similar to that of L* galaxies, but note that the result (\ref{dmin}) is fairly insensitive to the combination $Nn_g$ in any case. 

Since the distance $D \simeq 95$~Mpc suggested for HLX--1 is bigger than the minimum distance $D_{\rm min}$, an SS433--like identification of HLX--1 is so far not implausible. Encouragingly, $D$ also does not exceed $D_{\rm min}$ by large factors, which would require a further stringent constraint on the system visibility in addition to beaming to prevent us seeing systems closer to $D_{\rm min}$. Of course there actually is an obvious extra constraint of this kind, but it is relatively mild -- the emission of HLX--1 varies in time. We consider this next.

\section{Light Curves and Geometry}

X--rays from HLX--1 are bright for a significant fraction of its  $\sim 1$~yr cycle. This makes it plausible that it should be found at a distance $D$ not vastly greater than the minimum distance $D_{\rm min}$. 
The source is apparently never undetectable, varying by a factor $\sim 60$ \citep{Servillat11}. We ascribe the variation to the same underlying process inferred for SS433, namely that the central disc funnels collimating the radiation (and any material jet) precess approximately periodically.  In SS433 we do not look down the funnels, and the X--rays are very weak, probably representing a small component from the jets themselves \citep[cf ][]{Begelman06}. Instead the precession is mainly seen in the radial velocity measurements of the red-- and blue--shifted $H\alpha$ lines tracing the jets. The precession period in SS433 appears to vary over a range of a few days around some mean value $\sim 163$~d.  

We assume that the collimation of the disc funnels is purely geometrical. Then 
the X--ray light curve of HLX--1 simply reflects the motion and inherent variation of the precessing funnel visible to us. Since we have worked out that the beam of HLX--1 is rather narrow ($\theta_b \simeq 1.6^{\circ}$), the significant X--ray duty cycle $d \sim 0.5$ must mean that the angle $\phi$ between the beam and precession axes is itself comparable to the beam angular size $\theta_b$. Indeed if $d = 1$, i.e. we always see X--rays from the precessing funnel, we must have $\phi < \theta_b$. If instead we assume there are epochs with no beamed X--rays (either zero X--ray flux, or a low--level  unbeamed flux), then $\phi >\theta_b$. A possible source of a relatively constant low--level flux is X--ray emission from the jets themselves. Simple geometry, assuming a circular beam, gives $\phi = 5.7^{\circ}, 8.3^{\circ}$ for duty cycles $d = 0.5, 0.33$ respectively for the bright--phase emission. By contrast, if the precession angle $\phi$ were as large as in SS433 ($\phi \simeq 20^{\circ}$), the duty cycle would be only $d \simeq 0.15$, assuming the same $\dot m$ and thus $b$. 

These estimates illustrate how easily even small changes in geometry can alter the X--ray light curves quite significantly, even with a fixed beam size. In fact we should { \it expect} variations in the light curves: the simple fact that the beams (and any jets) move at all means that they cannot be anchored in structures with high inertia, such as the black hole spin axis or the innermost disc plane, which is closely tied to this spin axis through the Lense--Thirring effect \citep{NixonKing13}. Instead they must result from gaseous structures within a disc deformed by a process such as radiation warping \citep{Pringle96} and so subject to constant variation. The difficulty of calculating such effects means we cannot give a simple argument why the precession angle in HLX--1 should be rather smaller (factors 2 -- 3) than in SS433.

An obvious physical cause of inherent beam size variations is the sensitivity of the beam opening angle to accretion rate ($\theta_b \propto b^{1/2} \propto 1/\dot m$).  A beam narrowed by an increased accretion rate would for example cause the steep rise in the X--ray light curve to be significantly delayed, as recently observed. If we crudely model this as a change in bright--phase X--ray duty cycle from $d \sim 0.33$ to $d \sim 0.25$ (a $\sim $ one--month delay) we find this requires a change in $\theta_b \propto b^{1/2}$ corresponding to a change of only a factor 1.36 in $\dot m$.

An interesting question is what causes the bright--phase X--ray emission to have a characteristic rapid rise and slower decay. We must be looking directly down the X--ray beam at maximum, suggesting that the rapid rise corresponds to the funnel axis aligning with the line of sight. If this sharp collimation was the only modulation of the X--rays, we would see just a square--wave light curve ($\phi > \theta_b$). A simple interpretation of the slower decay we actually observe is that some scattering structure, perhaps a jet, gradually invades the line of sight as the beam precesses. A small offset of the jet from the beam axis is reasonable, since the reason for the jet precession (and its relatively low velocity) in SS433 is that a jet launched in a fixed direction (the black--hole spin axis) is deflected by a precessing gas structure \citep{Begelman06}.

We note that all the considerations above are purely geometrical. As a result they say nothing about the X--ray spectra or state changes seen in HLX--1. However the interpretation of the system as one with a highly super--Eddington mass transfer rate means that the accretion disc structure here is very different from that normally considered. There must be outflow of almost all of the transferred mass from all disc radii $R \la \dot m R_s \sim 100R_s$, where $R_s$ is the black hole Schwarzschild radius. A disc like this is well represented by a slim--disc model \citep{Veira14}.

\section{Discussion}

We have suggested that HLX--1 may be a stellar--mass binary system like SS433, but seen along its X--ray beams. The precession of these beams gives the $\sim 1$~yr characteristic timescale of the light curve, while the significant X--ray duty cycle means that the precession angle must be comparable with the beam opening angle. As a consistency check, the distance 
$\sim 95$~Mpc suggested for HLX--1 is only a few times larger than the minimum distance giving a reasonable chance of seeing such a highly--beamed system.

Put another way, we know that SS433 is simply a fairly normal high--mass binary in a very short--lived phase of its evolution, defined by the condition that the expanding blue supergiant companion is more massive than the compact accretor and currently fills its Roche lobe. Mass is then transferred on the thermal timescale of the supergiant. This phase automatically follows the high--mass X--ray binary (HMXB) phase. Any galaxy with recent star formation can host systems like this -- the recently--discovered system MQ1 in the galaxy M83 \citep{Soria14} may reveal another one \citep{King14}. So we would expect eventually to see an example with its radiation beam pointing towards us if we search a large enough sample of galaxies, that is, out to a sufficient distance ($D_{\rm min}$).  Since SS433 has one of the highest mass transfer rates likely to be observable in a stellar--mass binary, systems like it are the best stellar--mass candidates for the most extreme ULXs. These must be the shortest--lived, most tightly beamed, and brightest of the ULX population (both apparently and intrinsically).

We should then ask what the less luminous ULXs represent. All those with red companion stars -- a large fraction (Middleton, private communication) are probably soft X--ray transients in the course of long--lasting disc outbursts  (the so--called `GRS 1915--like' systems of King, 2002). ULXs with blue companions are probably instead systems which are either evolving towards the 
SS433/HLX--1 state or have just left it. The first group are moving from the wind--capture accretion of the HMXB phase towards the SS433/HLX--1 state, established once the nuclear expansion of the high--mass companion makes it fill its Roche lobe. The second group may have recently reversed their binary mass ratios through mass loss and transfer, so that the mass transfer process now stabilizes at a lower level than in the SS433/HLX--1 state. The MQ1 system referred to above may be a system like this \citep{King14}.

 \section*{Acknowledgments}

We thank Guillaume Dubus for stimulating discussions. JPL acknowledges support from the French Space Agency CNES and Polish NCN grants UMO-2011/01/B/ST9/05439, UMO-2011/01/B/ST9/05437 and DEC-2012/04/A/ST9/00083. ARK thanks several participants of the 2014 Leiden Lorentz Center workshop on ULXs, including Matt Middleton, Tom Maccarone, Peter Jonker and Cole Miller, for informative discussions. Theoretical astrophysics research at the University of Leicester is supported by an STFC Consolidated Grant.

\bsp

\label{lastpage}

\end{document}